\documentstyle[12pt]{article}

\def\greaterthansquiggle{\raise.3ex\hbox{$>$\kern-.75em\lower1ex\hbox{$\sim$}}}
\def\lessthansquiggle{\raise.3ex\hbox{$<$\kern-.75em\lower1ex\hbox{$\sim$}}}
\newcommand{\beq}{\begin{equation}}

\newcommand{\eeq}{\end{equation}}
\newcommand{\beqa}{\begin{eqnarray}}
\newcommand{\eeqa}{\end{eqnarray}}
\newcommand{\beqan}{\begin{eqnarray*}}
\newcommand{\eeqan}{\end{eqnarray*}}
\newcommand{\ba}{\begin{array}}
\newcommand{\ea}{\end{array}}
\newcommand{\no}{\nonumber}

\newcommand{\ve}{\varepsilon}

\newcommand{\wt}{\widetilde}
\newcommand{\wh}{\widehat}

\newcommand{\cL}{{\cal L}}

\newcommand{\cO}{{\cal O}}

\newcommand{\R}{{\cal R}}

\newcommand{\BV}{{\Bbb V}}

\def\nz{\ifmmode {I\hskip -3pt N} \else {\hbox {$I\hskip -3pt N$}}\fi}
\def\zz{\ifmmode {Z\hskip -4.8pt Z} \else
       {\hbox {$Z\hskip -4.8pt Z$}}\fi}
\def\qz{\ifmmode {Q\hskip -5.0pt\vrule height6.0pt depth 0pt
       \hskip 6pt} \else {\hbox
       {$Q\hskip -5.0pt\vrule height6.0pt depth 0pt\hskip 6pt$}}\fi}
\def\rz{\ifmmode {I\hskip -3pt R} \else {\hbox {$I\hskip -3pt R$}}\fi}
\def\cz{\ifmmode {C\hskip -4.8pt\vrule height5.8pt\hskip 6.3pt} \else
       {\hbox {$C\hskip -4.8pt\vrule height5.8pt\hskip 6.3pt$}}\fi}
\def\au{{\setbox0=\hbox{\lower1.36775ex%
\hbox{''}\kern-.05em}\dp0=.36775ex\hskip0pt\box0}}
\def\ao{{}\kern-.10em\hbox{``}}

{\catcode `\@=11 \global\let\AddToReset=\@addtoreset}
\AddToReset{equation}{section}

\newcommand{\subjclass}[1]{}

\def\scri{\hbox{${\cal J}$\kern -.645em {\raise
      .57ex\hbox{$\scriptscriptstyle (\ $}}}}

\newcommand{\commentout}[1]{}

\newcommand{\U}{{\cal U}}

\newcommand{\ee}{\end{equation}} \newcommand{\bea}{\begin{eqnarray}}
\newcommand{\eea}{\end{eqnarray}}
\newcommand{\beaa}{\begin{eqnarray*}}
\newcommand{\eeaa}{\end{eqnarray*}}

\newcommand{\Bbb}{\bf}
\begin{document} 

\title{Killing Initial Data}

\author{Robert Beig\thanks{Supported by Fonds zur F\"orderung der
wissenschaftlichen Forschung, Project P9376--PHY. {\em E--mail}: 
Beig@Pap.UniVie.AC.AT}
  \\Institut f\"ur Theoretische Physik\\ Universit\"at Wien\\ A--1090
  Wien, Austria\\ \\ Piotr T.\ Chru\'sciel\thanks{ 
 On leave of absence from the Institute of
    Mathematics, Polish Academy of Sciences, Warsaw.
  Supported in part by KBN grant \# 2P30105007,
 by the Humboldt Foundation
 and by the Federal Ministry of Science and Research, Austria.
   {\em E--mail}:
    Chrusciel@Univ-Tours.fr} 
\\ D\'epartement de Math\'ematiques\\Facult\'e
des Sciences\\ Parc de Grandmont\\ F-37200 Tours, France}

\maketitle

\begin{center}
We dedicate this work to Professor Andrzej Trautman on the occasion
of his birthday.
It is a great pleasure to pay tribute to his lasting contributions
to Relativity.
\end{center}
\begin{abstract} 
We study space--time Killing vectors in terms of their ``lapse and
shift" relative to some spacelike slice. We give a necessary and
sufficient condition  in order
for these lapse-shift pairs, which we call Killing initial data
(KID's), to form a Lie algebra under the bracket operation induced
by the Lie commutator of vector fields on space--time. This result
is applied to obtain a theorem on the periodicity of orbits for a class
 of Killing vectors fields in asymptotically flat space--times.
\end{abstract}

%\newpage
\section{Introduction}
\label{introduction}
When considering black hole space--times with more than one Killing
vector field it is customary to assume that one of the Killing vectors
has complete periodic orbits. In a recent paper [1] we have shown that 
this is necessarily the case, under a set of conditions on the
space--times under consideration. This set of hypotheses includes a
``largeness condition'' on the space--times, namely that the space--time
contains a ``boost-type domain''. While this hypothesis will be satisfied 
for many models of matter coupled to gravity, provided the fields
under consideration fall off sufficiently fast at spatial infinity
[2,3], there are various cases in which we are not {\em a priori\/}
certain that this will be the case. For this reason it is useful
to have results under hypotheses involving initial data sets only, and
that with a minimal set of hypotheses on the matter fields under
consideration. It is the aim of this paper to prove the existence of a
Killing vector with periodic orbits in a Cauchy data setting, when there
are at least two linearly independent Killing vectors, one of which is
transverse to the initial data surface (at least in the asymptotic
region). The reader should note that the
classification of possible isometry groups, or of possible Lie algebras
of Killing vectors, follows immediately from this result, as in [1],
except for one--dimensional algebras of Killing vectors.

In order to address the issue raised above, it is first necessary to
face the following problem: consider an initial data set with two or
more ``candidate Killing vector fields''. Under which conditions do
these vector fields lead to Killing vector fields on a corresponding
space--time? We show that this question can be reduced to that of
certain properties of an appropriately defined bracket operation on
the initial data surface. More precisely, we give a necessary and
sufficient condition for the bracket operation to form a Lie algebra.
We show that, when the bracket operation forms a Lie algebra, the
``candidate Killing vectors'' become Killing vectors in the Killing
development\footnote{See [4] and Sect. 2 of this paper for the 
definition of the notion of
Killing development. Let us emphasize that, when suitable field 
equations are imposed, there exists a neighbourhood of the initial
data hypersurface in the Killing development which is isometrically
diffeomorphic to a neighbourhood of the initial data surface in the
space--time obtained by evolving the inital data using the field
equations. Thus statements about the Killing developments are also
statements about solutions of the field equations, in this sense.}
associated with any ``transverse candidate Killing vector''.

This paper is organized as follows: in the next section we introduce
the notion of a ``Killing initial data'' (KID) and discuss some
elementary properties thereof. In Section 3 we give a sufficient and 
necessary condition for the set of KID's to form a Lie algebra.
In Section 4 we show that Lie algebras of KID's ``extend'' to Lie
algebras of Killing vectors of Killing developments. In Section 5
we consider asymptotically flat Killing developments of initial data
sets with at least 2 dimensional Lie algebras of KID's, and we prove the
existence of Killing vectors with periodic orbits in such a case.

\section{Killing initial data (KID's)}
Let $(M,g_{\mu\nu})$ be a connected spacetime and $X,\bar X$ be Killing
vector fields, i.e.
\beq
\cL_X g_{\mu\nu} = 0 = \cL_{\bar X} g_{\mu\nu}.
\eeq
Then the commutator $[X,\bar X]$ is also a Killing vector field since
\beq
\cL_{[X,\bar X]} g_{\mu\nu} = [\cL_X, \cL_{\bar X}] g_{\mu\nu} = 0.
\eeq
More generally, let {$\BV$} be the finite-dimensional vector space over
{$\Bbb R$} of Killing vector fields on $(M,g_{\mu\nu})$. Then {$\BV$} is
closed under $[\;,\;]$. Let $(\Sigma,g_{ij},K_{ij})$ be a connected
spacelike submanifold of $(M,g_{\mu\nu})$ with induced metric 
$g_{ij}$ and second fundamental form $K_{ij}$. We can then decompose
the Killing vector field $X$ along $\Sigma$ according to
\beq
X = N n^\mu \partial_\mu + Y^i \partial_i, \qquad
N = - X^\mu n_\mu
\eeq
where $n^\mu$ is the future unit normal of $\Sigma$. Here we are using
a coordinate system $x^\mu$ in which $\Sigma$ is described by the 
equation $t\equiv x^0 = 0$. In order to
translate the Killing equation into a statement in terms of $(N,Y^i)$ and
$(g_{ij},K_{ij})$ it is convenient to choose Gaussian coordinates
$x^\mu = (t,x^i)$ on a tubular neighbourhood of $\Sigma$ in $(M,g_{\mu\nu})$. 
%with $\Sigma$ given by $t = 0$. 
Then
\beq
g_{\mu\nu} dx^\mu dx^\nu = - dt^2 + g_{ij}(t,x^\ell)dx^i dx^j.
\eeq
The $(i,j)$-component of
\beq
\cL_X g_{\mu\nu} = X^\rho \partial_\rho g_{\mu\nu} + 2g_{\rho(\mu}
\partial_{\nu)} X^\rho = 0\ ,
\eeq
yields
\beq
2N K_{ij} + 2D_{(i}Y_{j)} = 0\ ,
\eeq
where we have used $\partial_t g_{ij} = 2K_{ij}$, valid in Gaussian
coordinates. The $(t,t)$-component of (2.5) says that
\beq
\partial_t N = 0
\eeq
and the $(t,i)$-component that
\beq
\partial_t Y^i = g^{ij} D_j N.
\eeq
Another interesting identity results from taking $\partial_t$ of
Equ. (2.6):
\beq
2N \partial_t K_{ij} + 2D_i D_j N + 4D_{(i} (K_{j)\ell} Y^\ell) -
2Y^\ell(2D_{(i} K_{j)\ell} - D_\ell K_{ij}) = 0
\eeq
where we have used (2.7,8). We now define
\beq
G_{\mu\nu} n^\mu n^\nu = \rho, \qquad
G_{\mu i} n^\mu = - J_i, \qquad
G_{ij} = \tau_{ij},
\eeq
where $G_{\mu\nu}$ is the Einstein tensor of $g_{\mu\nu}$.
The quantities $\rho$ and $J_i$ can be expressed in terms of $g_{ij}$
and $K_{ij}$ by the relations
\beqa
2\rho &=& {}^3R + K^2 - K_{ij} K^{ij} \\
- J_i &=& D^j(K_{ij} - K g_{ij}).
\eeqa
Using the well-known form of $G_{ij}$ in Gaussian coordinates to
eliminate $\partial_t K_{ij}$ from (2.9), we obtain
\beq
\cL_Y K_{ij} + D_i D_j N = N({}^3 \R_{ij} + K K_{ij} - 2K_{i\ell}K_j{}^\ell)
- N \left[ \tau_{ij} - \frac{1}{2} g_{ij} (\tau-\rho)\right],
\eeq
where $\tau := g^{ij} \tau_{ij}$. Clearly, Equ.'s (2.6) and (2.11--2.13)
hold independently of Gaussian coordinates.

In the Cauchy problem context it is often convenient to forget
about the space--time and consider only three dimensional initial data sets
$(\Sigma, g_{ij}, K_{ij})$. For the purpose of Equ. (2.13) we also
need to have  a tensor field
$\tau_{ij}$ defined on $\Sigma$. We shall call a pair
$(N,Y^i)$ a Killing initial data (KID), provided (2.6) and (2.13)
hold.

It is worthwile to point out that,  
in the context of the Einstein equations, $\tau_{ij}$ can be typically
calculated from the initial data  for the matter fields present, and is the
matter stress tensor.
An alternative point of view is the following: 
Consider a data set $(\Sigma, g_{ij}, K_{ij})$ together with
a scalar field $N$ and a 
vector field $Y^i$ satisfying Equ. (2.6). We can then use Equ. (2.11) to 
define a scalar field 
$\rho$, and then Equ. (2.13) to define a tensor field $\tau_{ij}$, at least
on the set where $N$ does not vanish. 
Thus if we have only one solution of Equ.        
(2.6), then  Equ. (2.13) is
trivial (except perhaps on the boundary of the zero set of $N$, if 
some regularity of $\tau_{ij}$ is imposed). If, however,  more than one pair
$(N,Y^i)$ solving  Equ. (2.6) exists, 
we can use one such solution to define
$\tau_{ij}$, and then consider only those solutions of Equ. (2.6) which
satisfy Equ. (2.13) with that given $\tau_{ij}$.

Given a KID on $(\Sigma,g_{ij},K_{ij})$, we can  ask the converse
question: Does there exist a spacetime $(M,g_{\mu\nu})$ ``evolving" from
$(\Sigma,g_{ij},K_{ij})$ with a Killing vector $X$ ``evolving" from
$(N,Y^i)$? There is an affirmative answer to this question  in the
following two cases:
\begin{description}
\item[Case 1:] $(N,Y)$ is ``transversal'', i.e., by definition,
 $N \neq 0$. Then we can 
use the KID $(N,Y^i)$ to define the {\em Killing development}
$(\wh M, \wh g_{\mu\nu})$ of $(\Sigma,g_{ij},K_{ij})$ (see [1]),
as follows: 
Let $\wh M = {{\Bbb R}} \times \Sigma$ and define the Lorentz metric
\beq
\wh g_{\mu\nu} dx^\mu dx^\nu = - \wh N^2 du^2 + \wh g_{ij}(dx^i +
\wh Y^i du)(dx^j + \wh Y^j du),
\eeq
$$
\wh N(u,x^i) = N(x^i), \qquad
\wh g_{ij}(u,x^\ell) = g_{ij}(x^\ell), \qquad
\wh Y^i(u,x^j) = Y^i(x^j).
$$
Then $\partial_u$ is a Killing vector of $(\wh M,\wh g_{\mu\nu})$ 
extending $(N,Y^i)$, that is: the vector field $X$ defined on $\Sigma$
by the right-hand-side of Equ. (2.3) coincides with the Killing vector
field $\partial_u$ there.
\item[Case 2:] $\rho = 0$, $J_i = 0$. In that case
when $g_{ij}$ and $K_{ij}$ are sufficiently regular,
 $(\Sigma,g_{ij},K_{ij})$ has
a vacuum 
Cauchy development $(\bar M, \bar g_{\mu\nu})$, i.e. $\bar R_{\mu\nu} = 0$.
If, furthermore, the KID $(N,Y^i)$ is a  vacuum KID in the sense
that the ``stress tensor'' $\tau_{ij}$, defined by Equ. (2.13) is
also zero, it is known (see [5] and references therein; cf. also [6]), that
the KID extends to a Killing vector on $(\bar M,\bar g_{\mu\nu})$.
\end{description}
An analogous statement holds when the vacuum
equation is modified by the presence of a cosmological constant
$\Lambda$, i.e. $\rho = - \Lambda$, $\tau_{ij} = \Lambda g_{ij}$,
$J_i = 0$.\\

Suppose, we now have a spacetime $(M, g_{\mu\nu})$ with two Killing 
vectors $X,\bar X$. Their commutator $[X,\bar X]$ gives rise, on
$(\Sigma,h_{ij},K_{ij})$, to the bracket
\beq
\{ (N,Y^i),(\bar N, \bar Y^j)\} := (\cL_Y \bar N - \cL_{\bar Y} N,
[Y,\bar Y]^\ell + N D^\ell \bar N - \bar N D^\ell N).
\eeq
This is the algebra  first studied in [7,8]. Note, however, that,
whereas in [7,8] the above bracket is, loosely speaking, a commutator of
vector fields in the infinite-dimensional space of spacelike
embeddings of some 3-manifold into spacetime, it arises in our case
simply from the commutator of Killing vector fields on spacetime.

We are now ready to ask the following question: Consider an
initial-data set
$(\Sigma,g_{ij},K_{ij})$ and two KID's, i.e. solutions of Equ. (2.6)
and Equ. (2.13) {\bf for the same} $\tau_{ij}$. Is their bracket,
defined by (2.15), also a KID with the same $\tau_{ij}$? An affirmative 
answer can immediately be given in the vacuum case (Case 2 above):
the vacuum development is clearly defined independently of the KID's,
and thus {\bf every} KID extends to a Killing vector field on
$(\bar M,\bar g_{\mu\nu})$. Thus the KID's, in this case, are closed
under $\{\;,\;\}$. In the non-vacuum case, when one of the KID's
$(N,Y^i)$ has $N \neq 0$, one might consider the Killing development
associated with this particular KID. But it is then unclear whether
some other KID $(\bar N, \bar Y^i)$, if present, extends to a
Killing vector in the Killing development given by $(N,Y^i)$. In fact,
the following example shows that KID's are in general {\bf not} closed
under $\{\;,\;\}$.
\paragraph{Example:} Let $(\Sigma,h_{ij},K_{ij}) = ({{\Bbb R}}^3,
\delta_{ij},0)$ and take for $\tau_{ij}$
\beq
\tau_{ij} dx^i dx^j = (dx^1)^2 + (dx^2)^2.
\eeq
Define two KID's by
\beq \ba{ll}
N= 0, \qquad & Y = x^2 \partial_{x^3} - x^3 \partial_{x^2} \\[6pt]
\bar N = e^{x^3}, \qquad & \bar Y = 0.\ea 
\eeq
It is then easy to check that $(N,Y^i)$ and $(\bar N, \bar Y^i)$ are
both KID's with $\tau_{ij}$ given by (2.16), but their bracket is not.

\section{The Lie algebra of KID's}
We first show the Jacobi identity for $\{\;,\;\}$.
\paragraph{Lemma:} Consider three pairs $(N,Y^i)$, $(\bar N, \bar Y^i)$, 
$(\wt N, \wt Y^i)$  satisfying Equ. (2.6). Then 
$$
\{(\wt N,\wt Y^i), \{ (N,Y^i),(\bar N, \bar Y^i)\}\} +
\{(\bar N, \bar Y^i), \{ (\wt N, \wt Y^i),(N,Y^i)\}\} +
$$
\beq
+ \{(N,Y^i), \{(\bar N,\bar Y^i),(\wt N,\wt Y^i)\}\} = 0.
\eeq

\paragraph{Proof:} This is a straightforward computation, based on the
Jacobi identity for the commutator of vector fields on $\Sigma$ and
relations like
\beq
\cL_Y D^i \bar N = D^i \cL_Y \bar N + 2 N K^{ij} D_j \bar N.
\eeq
\hfill\ $\Box$

We now state the main result of this paper.
\paragraph{Theorem:} Let {$\Bbb W$} be the vector space over {$\Bbb R$} of
KID's on $(\Sigma,g_{ij},K_{ij})$ for some fixed stress tensor
$\tau_{ij}$. The linear space {$\Bbb W$} is closed under the bracket
$\{\;,\;\}$, if and only if
\beq
(N \cL_{\bar Y} - \bar N \cL_Y) \tau_{ij} = 2J_{(i}(N D_{j)} \bar N
- \bar N D_{j)} N)
\eeq
for all pairs $(N,Y^i)$, $(\bar N,\bar Y^i)$ of KID's.
\paragraph{Proof:} We first have to look at the expression
\beq
\cL_{[Y,\bar Y] + ND\bar N - \bar N DN} g_{ij} + 
2(\cL_Y \bar N - \cL_{\bar Y} N) K_{ij},
\eeq
where $ND\bar N - \bar N DN$ is short-hand for the vector
$N D^i \bar N - \bar N D^i N$. Using Equ. (2.6) for both pairs
$(N,Y^i)$ and $(\bar N,\bar Y^i)$ the expression (3.4) can be written
as
\beq
\cL_Y(- 2 \bar N K_{ij}) + 2 D_{(i}(N D_{j)} \bar N) + 2 (\cL_Y \bar N) K_{ij}
- ((N,Y) \longleftrightarrow (\bar N,\bar Y)).
\eeq
Using Equ. (2.13) to eliminate $D_i D_j N$ and $D_i D_j \bar N$ in
(3.5), we find that all terms add up to zero. 
We are here, and in the following repeatedly, using that terms which are
independent of $Y$ and $\bar Y$ and
contain $N$ and $\bar N$ without derivatives drop out upon
antisymmetrization.
Thus $\{(N,Y),(\bar N,\bar Y)\}$ also satisfies Equ. (2.6). We now
compute $\cL_Y\; {}^3\R_{ij} = \delta\;{}^3\R_{ij}(\cL_Y g_{k\ell})$, 
where $\delta\;{}^3 \R_{ij}$ is the linearization of the Ricci tensor at
$g_{ij}$. Thus
\beqa
\cL_Y \; {}^3\R_{ij} &=& \Delta(N K_{ij}) + D_i D_j(NK) - \no \\
&& \mbox{} - 2 D_{(i} D^\ell (N K_{j)\ell}) - 2 N\;
{}^3\R^\ell{}_{(ij)}{}^m K_{\ell m} - \no \\
&& \mbox{} - 2 N\; {}^3\R_{(i}{}^\ell K_{j)\ell}.
\eeqa
Consequently
\beqa
(\bar N \cL_Y - N \cL_{\bar Y}){}^3\R_{ij} &=& 
\bar N[(\Delta N)K_{ij} + 2(D^\ell N)D_\ell K_{ij} + \no \\
&& \mbox{} + (D_i D_j N)K + 2(D_{(i}N)D_{j)}K - \no \\
&& \mbox{} - 2(D_{(i} D^\ell N)K_{j)\ell} - 2(D^\ell N)
D_{(i} K_{j)\ell} - \no \\
&& \mbox{} -2(D_{(i}N) (D_{j)} K - J_j)] - (N \longleftrightarrow
\bar N) = \no \\
&=& \bar N[(\Delta N)K_{ij} - (\cL_Y K_{ij})K + 
2(\cL_Y K_{(i}{}^\ell)K_{j)\ell} + \no \\
&& \mbox{} + 2D^\ell N(D_\ell K_{ij} - D_{(i}K_{j)\ell}) + 
2(D_{(i}N) J_{j)}] - \no \\
&& \mbox{} - ((N,Y)  \longleftrightarrow (\bar N, \bar Y))
\eeqa
where we have used (2.13) in the last line. Equ. (3.6) 
and (2.13) imply that
\beq
\cL_Y K = N({}^3\R + K^2) - \Delta N - N \left( - \frac{\tau}{2} +
\frac{3}{2} \rho \right).
\eeq
Now Equ.'s (3.7) and (3.8) and the definition (2.11) of $\rho$ give
rise to
\beq
(N \cL_{\bar Y} - \bar N \cL_Y)\rho = 2(N D^i \bar N - \bar N D^i N)J_i.
\eeq
We finally have to compute
\beq
\cL_{[Y,\bar Y]+ND\bar N-\bar NDN} K_{ij} + D_i D_j(\cL_Y \bar N -
 \cL_{\bar Y}N) - (\cL_Y \bar N - \cL_{\bar Y} N) M_{ij}
\eeq
where $N M_{ij}$ is the r.h.side of (2.13), i.e.
\beq
M_{ij} := {}^3\R_{ij} + K K_{ij} - 2K_{i\ell} K_j{}^\ell -
\tau_{ij} + \frac{1}{2} g_{ij}(\tau - \rho).
\eeq
Using (2.13), the expression (3.10) turns into
\beqa
&& \mbox{} - [\cL_Y,D_iD_j]\bar N + \bar N \cL_Y M_{ij} + 
N(D^\ell \bar N)D_\ell K_{ij} + \no \\
&& \mbox{} + 2K_{\ell(i} D_{j)} (N D^\ell \bar N) -
((N,Y) \longleftrightarrow (\bar N,\bar Y)) = \no \\
&=& \bar N \cL_Y M_{ij} - 2 \bar N(D^\ell N)(D_\ell K_{ij} - D_{(i}
K_{j)\ell}) + \no \\
&& \mbox{} + 2 \bar N K_{\ell(i} (\cL_Y K_{j)}{}^\ell) -
((N,Y) \longleftrightarrow (\bar N,\bar Y)).
\eeqa
We now insert Equ.'s (3.7,8) into $(\bar N \cL_Y - N \cL_{\bar Y})M_{ij}$
and substitute this in the third line of (3.12). Remarkably, all terms not 
involving $\tau_{ij}, J_i,\rho$ drop out. In order for
$\{(N,Y),(\bar N,\bar Y)\}$ to again satisfy Equ. (2.13), we are then
left with the condition
\beq
(N \cL_{\bar Y} - \bar N \cL_Y)\tau_{ij} - \frac{1}{2} g_{ij}
(N \cL_{\bar Y} - \bar N \cL_Y)(\tau - \rho) =
2 J_{(i} (N D_{j)} \bar N - \bar N D_{j)} N).
\eeq
%Taking the trace of (3.13) and using Equ. (3.9), we find that the
%second term on the left in (3.13) is zero. 
It is easily seen from (3.9) that (3.13) is equivalent to (3.3).
Thus we are left with 
(3.3) as the necessary and sufficient condition for {$\Bbb W$} to 
form a Lie algebra under $\{\;,\;\}$, and the proof is complete.
\hfill\ $\Box$

We also record, for later use, the identity
\beq
(N \cL_{\bar Y} - \bar N \cL_Y)J_i =
(N D^j \bar N - \bar N D^j N)\tau_{ij} + 
(N D_i \bar N - \bar N D_i N)\rho.
\eeq
Equ. (3.14) follows from the definition (2.12) and Equ.'s (2.6,13), 
independently
of the condition (3.3), in much the same way as (3.9) follows from
(2.11). 

There are situations, in addition to the vacuum case, where the
condition (3.3) is ``automatically satisfied''. Let $\rho$ be
everywhere positive and suppose that
\beq
\tau_{ij} = \frac{1}{\rho} J_i J_j.
\eeq
Then (3.3) follows from (3.9) and (3.14). If there is a transversal
KID $(N,Y^i)$ and if, in addition to (3.15), there holds
\beq
\rho = \sqrt{J_i J^i} ,
\eeq
the Killing development associated with $(N,Y^i)$ is a null dust
spacetime, i.e.
\beq
\wh G_{\mu\nu} = \wh \rho \xi_\mu \xi_\nu , \qquad
\wh g^{\mu\nu} \xi_\mu \xi_\nu = 0
\eeq
with $\wh \rho(u,x^i)=\rho(x^i)$,  $\wh J_\ell(u,x^i)=J_\ell(x^i)$, and
\beq
\xi_\mu dx^\mu = \wh N du - \frac{1}{\wh \rho} \wh J_i(dx^i + 
\wh Y^i du).
\eeq
Another possibility would be to have $\rho \ge 0$ (and not necessarily
identically vanishing), $\tau_{ij} = 0$, $J_i = 0$.
Then any Killing development is a (standard, i.e. non-null) dust spacetime.

Finally, there is the situation where $(\rho,J_i,\tau_{ij})$ are built
from some other (``good matter'') fields, i.e. fields with the property
that the combined Einstein-matter system allows a properly posed
initial-value problem. For example, $(\rho,J_i,\tau_{ij})$ could be
built from the $(E_i,B_i)$-fields derived from a Maxwell field
$F_{\mu\nu}$. Then, when there is a spacetime Killing field $X$
satisfying in addition that $\cL_X F_{\mu\nu} = 0$, the KID associated
with $X$ would satisfy some further equations involving $(E_i,B_i)$.
This is discussed in more detail in Section 5.
Conversely (see [9]) any KID satisfying these latter equations extends
to a Killing vector on the Einstein-Maxwell spacetime evolving from
$(\Sigma,g_{ij},K_{ij};E_i,B_i)$. Thus the condition (3.3) is
again automatically satisfied in this case, when $E_i$ and $B_i$ are
invariant in an appropriate sense, cf.\ eq.\ (5.2) below.

\section{Killing developments}
Suppose now that condition (3.3) is valid and we have a (nontrivial)
Lie algebra of KID's. Suppose, further, that $(N,Y^i)$, one of these
KID's, has $N \neq 0$, so that we can consider the Killing development
defined by $(N,Y^i)$. Then we have
\paragraph{Proposition:} Consider an initial data set $(\Sigma,g_{ij},
K_{ij})$ and suppose that the set of KID's forms a Lie algebra {$\Bbb W$}.
Assume further that there exists a KID $(N,Y^i)$ in {$\Bbb W$} such that
$N > 0$, and denote by $(M,g_{\mu\nu})$ the Killing development of
$(\Sigma,g_{ij},K_{ij})$ based on $(N,Y^i)$. Then there is a one-to-one
correspondence between the Killing vectors of $(M,g_{\mu\nu})$ and KID's,
which preserves the Lie algebra structure of {$\Bbb W$}.

\paragraph{Proof:} In the Killing development of $(N,Y^i)$, the extension
$\wh X$ of $(N,Y^i)$ is given by
\beq
\wh X = \wh N n^\mu \partial_\mu + \wh Y^i \partial_i = \partial_u
\eeq
when $u_\mu$ is the unit future normal to $u =$ constant. When
$(N_\alpha,Y^i_\alpha)$ is any other KID we have by assumption that
\beq
\{(N,Y),(N_\alpha,Y_\alpha)\} = c_\alpha(N,Y) + c_\alpha{}^\beta
(N_\beta,Y_\beta)
\eeq
for some constants $c_\alpha,c_\alpha{}^\beta$. We now define extensions 
$\wh X_\alpha$ of these KID's by the system of linear homogeneous
ODE's
\beqa
\partial_u \wh N_\alpha &=& c_\alpha \wh N + c_\alpha{}^\beta 
\wh N_\beta \no \\
\partial_u \wh Y_\alpha{}^i &=& c_\alpha \wh Y^i + c_\alpha{}^\beta 
\wh Y_\beta{}^i
\eeqa
with $\wh N_\alpha(0,x^i) = N_\alpha(x^i)$, $\wh Y_\alpha{}^i(0,x^j) =
Y_\alpha{}^i(x^j)$ and
\beqa
\wh X_\alpha &=& \wh N_\alpha n^\mu \partial_\mu + \wh Y_\alpha{}^i
\partial_i = \\
&=& \frac{1}{\wh N} \wh N_\alpha \partial_u + 
\left( \wh Y_\alpha{}^i - \frac{\wh N_\alpha}{\wh N} \wh Y^i \right)
\partial_i.
\eeqa
We now compute $\cL_{\wh X_\alpha} \wh g^{\mu\nu}$ for $u = 0$ with
$\wh g_{\mu\nu}$ given by Equ. (2.14), i.e.
\beq
\wh g^{\mu\nu} \partial_\mu \partial_\nu = - \frac{1}{\wh N^2}
(\partial_u - \wh Y^i \partial_i)(\partial_u - \wh Y^j \partial_j) +
\wh g^{ij} \partial_i \partial_j.
\eeq
We find that the $(uu)$-component of
$\cL_{\wh X_\alpha} \wh g^{\mu\nu}$ vanishes by virtue of
\beq
\partial_u \wh N_\alpha = \cL_Y N_\alpha - \cL_{\bar Y_\alpha} N,
\eeq
which follows from (2.15) and (4.2,3). Furthermore the $(ui)$-components vanish
by virtue of
\beq
\left. \partial_u \wh Y_\alpha{}^i\right|_{u=0} = [Y,Y_\alpha]^i +
N D^i N_\alpha - N_\alpha D^i N.
\eeq
Finally the $(ij)$-component of $\cL_{\wh X_\alpha} \wh g^{\mu\nu}$
is zero for $u = 0$, by virtue of $(N,Y)$, $(N_\alpha,Y_\alpha)$ all
obeying Equ. (2.6) (Equ. (2.6) actually coincides with
the $(ij)$-component of $\cL_{\wh X_\alpha} \wh g_{\mu\nu}=0$). Furthermore 
we see from (4.5) and $\wh X = \partial_u$ that
\beq
[\wh X,\wh X_\alpha] = c_\alpha \wh X + c_\alpha{}^\beta \wh X_\beta
\eeq
for all $u \in {{\Bbb R}}$. Thus
\beq
\frac{\partial}{\partial u} \left( \cL_{\wh X_\alpha} \wh g_{\mu\nu}\right)
= c_\alpha{}^\beta \cL_{\wh X_\beta} \wh g_{\mu\nu},
\eeq
which, combined with $\left. (\cL_{\wh X_\alpha} \wh g^{\mu\nu})
\right|_{u=0} = 0$, gives the result that $\wh X_\alpha$ is a Killing 
vector of $\wh g_{\mu\nu}$, as required. \hfill\ $\Box$

We can now interpret the meaning of the condition (3.3) in terms of
Killing developments. Suppose $X$ is a Killing vector of $(M,g_{\mu\nu})$
with complete orbits, intersecting exactly once an everywhere transversal 
spacelike
submanifold with induced metric $g_{ij}$. It follows that there exist
coordinates $(u,x^i)$, $- \infty < u < \infty$, such that
$$
g_{\mu\nu} dx^\mu dx^\nu = - N^2 du^2 + g_{ij}(dx^i + Y^idu)(dx^j +
Y^j du)
$$
with $N,Y^i$ and $g_{ij}$ all independent of $u$ and $X = N n^\mu 
\partial_\mu + Y^i \partial_i = \partial_u$. Suppose there exists
another Killing vector 
\beq
\bar X = \bar N n^\mu \partial_\mu + \bar Y^i \partial_i 
= \frac{\bar N}{N} (\partial_u - Y^i \partial_i) + \bar Y^i 
\partial_i.
\eeq
By the $(uu)$- and $(ui)$-components of $\cL_{\bar X} g^{\mu\nu} = 0$
we find that
\beqa
\partial_u \bar N &=& \cL_Y \bar N - \cL_{\bar Y} N \no \\
\partial_u \bar Y^i &=& [Y,\bar Y]^i + N D^i \bar N - \bar N D^i N.
\eeqa
We also know that
\beq
\cL_{\bar X}  G_{\mu\nu} = 0,
\eeq
where $G_{\mu\nu}$ is the Einstein tensor of $g_{\mu\nu}$. Writing
this out, using (4.11,12) and  
\beq
G_{\mu\nu} dx^\mu dx^\nu = N^2 \rho du^2 + [\tau_{ij}(dx^i + Y^idu) -
2N J_j du](dx^j + Y^j du),
\eeq
we find after straightforward manipulations that (4.13) is equivalent
to Equ.'s (3.3), (3.9) and (3.14). Since (3.9) and (3.14) are just
identities, we have thus found that (3.3) is merely the condition for
${\bar X}$ defined by (4.11) to be a vector field in the Killing
development $(M,g_{\mu\nu})$ of $(N,Y)$ which Lie derives
$G_{\mu\nu}$. We can now ask whether (4.13) is already sufficient for
${\bar X}$ to be a Killing vector of $g_{\mu\nu}$.  
In other words: Suppose we have a transversal KID $(N,Y)$. Define
$\tau_{ij}$ by (2.13). Suppose further we have another KID
$(\bar N,\bar Y)$ compatible with $(N,Y)$ in that it satisfies (2.13)
{\bf with the same} $\tau_{ij}$, and $\tau_{ij}$ satisfies (3.3). Then: is $\bar X$, defined by
\beq
\bar X^\mu(u,x)\partial_\mu = \frac{\bar N(u,x)}{N(x)}
(\partial_u - Y^i(x) \partial_i) + \bar Y^i(u,x) \partial_i
\eeq
with $\bar N(u,x),\bar Y^i(u,x)$ obeying Equ.'s (4.12), a Killing
vector of $(M,g_{\mu\nu})$? We believe the answer in
general will be no, for the following reason: 
Suppose $(\Sigma,g_{ij},K_{ij})$, $(N,Y)$, $(\bar N,\bar Y)$ are all
analytic for $u = 0$. Then, by the Cauchy--Kowalewskaja theorem, the 
evolution equations (4.12) can be solved for $(\bar N,\bar Y)$, whence
4 components of $\cL_{\bar X} g^{\mu\nu} = 0$ are already satisfied.
Equ. (2.6) however is a priori only valid for $u = 0$. The condition
for the $u$-derivative of this equation to vanish is precisely Equ.
(3.3). This, in turn, is again only valid for $u = 0$, and there is no
guarantee that Equ. (3.3) will propagate. The previous Proposition
circumvents this problem by assuming that (3.3) be satisfied for all
pairs of KID's with the same tensor field $\tau_{ij}$. 
In the case where $(\rho,J_i,\tau_{ij})$ is built from ``good matter
fields'', the Killing development will, in the domain of dependence of
$\Sigma$, be the same as the solution to the coupled system, in which
case the propagation of Equ. (3.3) is automatically taken care of.

\section{An application: Periodicity of Killing orbits}

A prerequisite for the classification of stationary black-holes is an
understanding of possible isometry groups of asymptotically flat
space--times. A classification of the latter has been recently established
in [1], under a ``largeness condition'' on the space--times under
consideration. As an application of our results in the preceding sections, 
we show below that the results of [1] can be recovered without
any space--time ``largeness'' conditions, when two or more Killing
vector fields are present, one of them being transverse to the initial 
data hypersurface $\Sigma$.

Consider, thus, as in the preceding section, an initial data set $(\Sigma,
g_{ij},K_{ij})$ with a KID $(N_0,Y_0^i)$, with $N_0 > 0$.
If one imposes well-behaved evolution equations on the metric, one
expects that in the resulting space--time $(\wh M,\wh g_{\mu\nu})$ there
will exist a neighbourhood $\cO \subset \wh M$ of $\Sigma$ and an
appropriate coordinate system on $\cO$ such that the metric will take
the form (2.14) (with $\wh N$ replaced by $\wh N_0$, etc.), with
$u \in (u^-(p),u^+(p))$, $p \in \Sigma$, $u^+ \in {{\Bbb R}}^+ \cup
\{\infty\}$, $u^- \in {{\Bbb R}}^- \cup \{-\infty\}$. (This will be the
case if e.g. the vacuum Einstein equations are imposed.) This provides
us with an isometric diffeomorphism $\Psi$ between $\cO$ and the subset
$\U = \{ p \in \Sigma, u^-(p) < u < u^+(p)\} \subset M$, where
$(M,g_{\mu\nu})$ denotes the Killing development of $(\Sigma,g_{ij},
K_{ij})$ based on $(N_0,Y_0^i)$.\footnote{Some results concerning the 
question, under which conditions $\U = \wh M$, $\cO = M$, can be found in
[10,11,12].} 
One can thus gain insight into the structure of the Killing orbits in
$\wh M$ by studying that of the Killing orbits of $M$: indeed, if the
orbit of a Killing vector field through a point $q \in \U$ {\bf always} 
remains in $\U$, then one will obtain a complete description of the 
corresponding orbit of the corresponding Killing vector field on $\wh M$.
We wish to point out the following result, which is a straightforward
consequence of the results of Section 4 and of [1,4].

\paragraph{Theorem:} Let $(\Sigma ,g_{ij},K_{ij})$ be 
an asymptotically
flat end in the sense of [4], i.e.
$ \Sigma\equiv\Sigma_R \equiv {\bf R}^3 \setminus B(R)$, $R > 0$ with
$(g_{ij},K_{ij})$ satisfying\footnote{ We write $f=O_k(r^{-\alpha})$
  if there exists a constant $C$ such that
  $|r^{-\alpha}f|+\ldots+|r^{-\alpha-k}\partial_{i_1\ldots i_k}f| \le C$.}
\beq
g_{ij} - \delta_{ij} = O_k(r^{-\alpha}), \qquad
K_{ij} = O_{k-1} (r^{-1-\alpha}),
\eeq
with $k \geq 3$ and $\alpha > 1/2$.
Let $|\rho| + |J^i| = O(r^{-3-\ve})$,
$\ve > 0$, and let the ADM four-momentum of $\Sigma$ be timelike. Consider
a tensor field $\tau_{ij}$ on $\Sigma$ satisfying  
$|\tau_{ij}| = O(r^{-3-\ve})$, and let {$\Bbb W$} denote the set
of solutions $(N,Y^i)$ of the equations (2.6) and (2.13),
%\beq
%2N K_{ij} + 2D_{(i} Y_{j)} = 0,
%\eeq
 suppose that {$\Bbb W$} is closed under the bracket (2.15). Assume
%further
%that the stress tensor satisfies $|\tau_{ij}| = O(r^{-3-\ve})$, and 
finally that
there exists $(N_0,Y_0^i) \in {\Bbb W}$ such that $N_0 > 0$, and 
let $(M,g_{\mu\nu})$
be the Killing development of $(\Sigma,g_{ij},K_{ij})$ based on
$(N_0,Y_0^i)$. Then for every $(N,Y^i) \in {\Bbb W}$ there exists a constant
$a \in {{\Bbb R}}$ such that the KID $(\wh N,\wh Y^i)$ defined as
$(N,Y) + a(N_0,Y_0^i)$ gives rise to a Killing vector on $(M,g_{\mu\nu})$
which has {\em complete periodic orbits},  through those
points $p$ in the asymptotically flat region for which $r(p)$ is
large enough. Moreover the set $\{ \wh N = \wh Y^i = 0\}$ is not empty.

\paragraph{Remarks:} 1. The Theorem proved in Section 3 gives a 
necessary and sufficient condition for  {$\Bbb W$} to be closed 
under the bracket $\{\cdot,\cdot\}$. This condition is trivially 
satisfied in vacuum ($\rho=J_i=\tau_{ij}=0$). As mentioned at the 
end of Section 3, it is 
also satisfied in electro-vacuum when the KID's correspond moreover to
``a symmetry'' of the initial data of the Maxwell field. 
More precisely, let $E_i,B_i$ be vector fields on $(\Sigma, g_{ij},
K_{ij})$ satisfying 
\beqa
\rho &=& \frac{1}{2} (E_i E^i + B_i B^i) \no \\
J_i &=& \ve_i{}^{jk} E_j B_k \equiv (E \times B)_i \no \\
\tau_{ij} &=& \frac{1}{2} g_{ij} \rho - (E_i E_j + B_i B_j) \\
\cL_Y E_i &=& NKE_i - 2NK_i{}^j E_j - N(D \times B)_i -
(DN \times B)_i \no \\
\cL_Y B_i &=& NKB_i - 2NK_i{}^j B_j + N(D \times E)_i + 
(DN \times E)_i. \no
\eeqa

These conditions arise as follows: Let $F_{\mu\nu} = F_{[\mu\nu]}$ be
a two-form on spacetime $(M,g_{\mu\nu})$ with $(\Sigma, g_{ij},
K_{ij})$ a spacelike submanifold and Killing vector $X = N n^\mu
\delta_\mu + Y^i \delta_i$. Write 
\beq
F_{\mu\nu} = 2 E_{[\mu} n_{\nu]} + \epsilon_{\mu\nu\rho\sigma} B^\rho
n^\sigma,  
\eeq
with $E_\mu n^\mu = 0 = B_\mu n^\mu$ and
$\epsilon_{\mu\nu\rho\sigma}\epsilon^{\mu\nu\rho\sigma} = - 24$, 
$\epsilon_{0123} > 0$. Then let 
\beq
G_{\mu\nu} = F_{\mu\rho} F_\nu{}^\rho - 1/4\,g_{\mu\nu} F_{\rho\sigma}
F^{\rho\sigma}. 
\eeq
Equ. (5.4) implies the first three conditions of (5.2). Now impose 
${\cal L}_X F_{\mu\nu}=0$ and
\beqa
\nabla^\mu F_{\mu\nu}& = & 0 \no \\ 
\nabla_{[\mu} F_{\nu\rho]} & = & 0.
\eeqa
Then
% the (ij)-component of 
the first of 
Equ.'s (5.5) implies the fourth of
(5.2) and 
%the (ij)-component of 
the second of Equ.'s (5.5) implies the fifth of (5.2). 

We claim that Equ.'s (5.2) imply Equ. (3.3). In checking that one can use
the following identity
\beq
A_i(B \times C)_j + C_i(A \times B)_j + B_i(C \times A)_j =
g_{ij} A^k(B \times C)_k,
\eeq
where $A_i,B_i,C_i$ are vector fields on $\Sigma$.

2. The condition that  $\{ \wh N = \wh Y^i = 0\}\ne 0$ is the usual
condition of axi-symmetry. This condition is needed e.g. to be able
to perform the  reduction of the axi-symmetric stationary 
electro--vacuum equations to the well known harmonic map equation.

\paragraph{Proof:} By the Proposition in Section 3 the Lie algebra of
Killing vector fields of $(M,g_{\mu\nu})$ is isomorphic to the Lie
algebra of KID's. The result is obtained by a repetition of the
arguments of [1], no details will be given. Let us simply point out
that the hypothesis of completeness of Killing orbits made in
Theorem 1.2 of [1] was done purely for the sake of simplicity of the
presentation of the results proved. \hfill\ $\Box$

\end{document}